\documentclass[12pt]{iopart}
\usepackage{epsfig}
\begin{document}

\title[Selective production of metallic carbon nanotubes]
{Selective production of metallic carbon nanotubes}

\author{Yasushi Matsunaga, Tadatsugu Hatori\dag\  and Tomokazu Kato}

\address{Graduate School of Science \& Engineering, Waseda University, 
Shinjuku-ku, Tokyo 169-8555, Japan}

\address{\dag Faculty of Science, Tsuchiya, Hiratsuka, Kanagawa 259-1293, Japan}

\date{today}

\begin{abstract}
  In this report, 
  we discuss whether the optimal electric field to promote the growth of armchair- type nanotubes (metallic character) evaluated 
   using the previous H$\ddot{\mbox{u}}$ckel-Poisson method can be applied at  the tip of a nanotube in a realistic system. 
Setting the cross-section of a nanotube and the external field by the sheath, 
  we estimate an effective area at the sheath edge. 
  Since the electric charge distribution in a nanotube caused by the external electric 
  field was determined  in our previous study, 
  we obtained the electric field distribution  out of  a nanotube 
  by solving the Poisson equation and clarified the structure of the electric field lines.
  By determining the effective area,  
  we show the optimal interval of the catalyst metal, 
  which is necessary to selectively grow the metallic nanotubes. 
  When nanotubes grow thickly during  the initial stage of growth, a strong electric field cannot be applied to the tips of the tubes. 
  As a tube grows and the tube length increases,  we found that the optimal electric field decreased. 
  To maintain the chemical activity at the tip, the sheath electric field must be decreased. 
  We estimated the decreasing rate of the sheath field to the tube length. 
  

\end{abstract}



\section{Introduction}
   It is known that SWCNT (single-walled carbon nanotube) 
   has the characteristics of a metal or semiconductor according to the chiral angle\cite{rf:iijima}.  
   However, the production method controlling the chiral angle has not been established. 
   The method of controlling the chiral angle by the electric field and selectively producing an armchair type (metallic character)  was then investigated in our previous study\cite{rf:ymatsunaga}. 
   The electronic states of the $\pi$ electrons are described by 
   the H$\ddot{\mbox{u}}$ckel method and their electric interactions are self-consistently 
   taken into account through the Poisson equation. 
   The frontier electron density at both ends of the nanotubes with open-ends was evaluated\cite{rf:fukui,rf:fukui2,kumeda1,kumeda2}. 
The electric field intensity to promote the growth of  the armchair-type nanotube was found. 
 The optimal intensity of  a direct current electric field to efficiently make metallic nanotubes is approximately 1V/nm\cite{rf:ymatsunaga}.

In this study, we examined whether  the imposition of  this optimal electric field on nanotubes through a sheath electric field
is possible,  
and evaluated the interval required to grow  a long nanotube. 
To selectively produce a long nanotube  by the electric field, 
it is clear that the restriction arises at the interval of the tubes on a substrate.  
This is because an effective electric field will not be applied at the tip of a nanotube 
if the interval of  the nanotubes is too narrow. 
Using the electric charge distribution of a nanotube from our previous result, 
and solving the distribution of the electric field out of a nanotube, 
we estimated the interval of nanotubes required to actually apply
 the optimal electric field evaluated in  our previous  study.

In section 2,  we present the calculation assumptions and some comments.  
The results are given in section 3.  The interval $R$ and the electric field lines are shown.  
In section 4,  the relation between the electric charge induced by the external field 
and the tube length is shown. 
We discuss the dependence of the sheath electric field on the tube length 
with the change in the induced charge. 
We summarize this study in section 5.

\section{Settings and calculation of electric field}

We assumed a single-walled nanotube (SWNT) and a constant sheath electric field. 
We approximate  the electric charge distribution of a nanotube 
by the distribution of the multi-rings. 
The axial electric charge distribution is given by our previous results 
using the H$\ddot{\mbox{u}}$ckel-Poisson method\cite{rf:ymatsunaga}. 
We approximate the electric field distribution out of the nanotube formed by multi-rings 
with the electric charge by the Legendre function.

If a sheath electric field completely concentrates at the tip of a nanotube, 
the relation of the electric field 
and the surface area can be estimated to be $E_0S_0=E_1 S_1$ based on   Gauss's law. 
Here, $E_0$ denotes the sheath electric field, $S_0$ denotes the area at  the sheath edge, $E_1$ denotes the electric field at the tip of a nanotube, and $S_1$ denotes  the cross section of a nanotube.   
However, 
 because the  charge induced in the nanotube influences the field lines 
and the electric field opposite to the sheath field occurs, 
$E_0S_0=E_1 S_1$ is not valid when the tube length is short as compared to the sheath length. 
 When an electric charge distribution of a nanotube  is  given,  we can calculate the area $S_{0} $ at a sheath edge.   
 If $S_{0} $ is found, 
 the interval of the metal catalyst patterned on a substrate 
 is able to be theoretically estimated. 
 That is, optimal numbers per unit area of nanotubes which can be effectively grown   is found. 
Moreover, the dependence of $S_{0} $  on the nanotube length  is evaluated. 
This is because the electric field on a tip of a tube changes with the tube length.
We use the following notations and parameters as shown in table \ref{table:1}. 
Figure 1 shows the outline of the electric field lines out of a nanotube.










We set  the origin of the coordinate ($r$,$\theta$,$\phi$) to the center of a nanotube, 
as shown in figure 1. 
From the symmetry of a tube, the electric field depends only on $r$ and $\theta$.   
We express the position of the $i$th-ring in the $z$ direction as $l_i$. 
The electric potential caused by the charge in the $i$th-ring is approximated by the Legendre function
 using the following equation,  
\begin{eqnarray}
\phi_i=\frac{M_i}{r_i}(1-\frac{1}{2}(\frac{a}{r_i})^2P_2(\cos\theta_i))+..). 
\end{eqnarray}
Here, we set $M_i=\frac{\lambda_i a}{2\epsilon_0}$, where $M_i$ denotes 
the $i$th ring charge divided by $2\epsilon_0$ 
and $\lambda_i$ denotes the $i$th charge per length.  
Also, $r_i$ and $\theta_i$ are given by
%
 \begin{eqnarray}
r_i=\sqrt{r^2+l^2_i-2 r l_i \cos\theta},
\\
\sin\theta_i=\frac{r\sin\theta}{\sqrt{r^2+l_i^2-2r l_i\cos\theta}}. 
\end{eqnarray}
The total potential $\phi$  is summed to $\phi_i$ using the following equation,   
\begin{eqnarray}
\phi(r, \cos\theta)=\sum_{i=1}^n \phi_i(r_i, \cos\theta_i). 
\end{eqnarray}

We calculate the electric field  lines and examine whether the electric field lines at the sheath edge 
arrive at the tip of a nanotube.  
We define $R=L \tan\theta_0$, then $R$ denotes the distance at the sheath edge 
from the nanotube axis as shown in figure 1.  
We examine $R$, 
by changing the tube length or the charge density in the nanotube.   
We set the initial condition  at the sheath edge and calculate the following equation.  
\begin{eqnarray}
\frac{dr}{d\theta}=\frac{rE_r}{E_{\theta}}. 
\end{eqnarray}
We write each component using the following equations, 
\begin{eqnarray}
E_r'=-\sum_i\frac{\partial \phi_i}{\partial r}=-\sum_i(\frac{\partial \phi_i}{\partial r_i}
\frac{\partial r_i}{\partial r}+\frac{\partial \phi_i}{\partial \theta_i}
\frac{\partial \theta_i}{\partial r}), 
\\
E_{\theta}'=-\frac{1}{r}\frac{\partial \phi}{\partial \theta}=
-\frac{1}{r}\sum_i\frac{\partial \phi_i}{\partial \theta}=-\frac{1}{r}\sum_i(\frac{\partial \phi_i}{\partial r_i}
\frac{\partial r_i}{\partial r}+\frac{\partial \phi_i}{\partial \theta_i}
\frac{\partial \theta_i}{\partial \theta}), 
\\
E_r=E_r'-E_0\cos\theta, 
\\
E_{\theta}=E_{\theta}'+E_0\sin\theta. 
\end{eqnarray}
$E_0$ denotes the sheath electric field. Each component is given by 
\begin{eqnarray}
\frac{\partial r_i}{\partial r}=\frac{r-l_i\cos\theta}{r_i},
\\
\frac{\partial r_i}{\partial \theta}=\frac{rl_i\sin\theta}{r_i},
\\
\frac{\partial \theta_i}{\partial r}=\frac{\pm l_i\sin\theta}{r_i^2},
\\
\frac{\partial \theta_i}{\partial \theta}=\frac{\pm r(r_i^2\cos\theta-rl_i\sin\theta^2)}
{r_i^2(r\cos\theta-l_i)}.
\end{eqnarray}

\section{Results}
First, we show the result for  two rings, as shown in figure 2. 
Then, $m$ denotes the averaged $\pi$ electron number per ring and  
the symbol $M$ is defined by $\frac{me}{4 \pi \epsilon_0}$.  
We dropped the $i$th suffix.   
The radius $a$ is set to 1 nm and $m$ is 2.3.  
The sheath electric field $E_0$ is set to 0.01 V/nm. 
The change of the distance $R$ with the change in the length $l$ is shown in table \ref{table:R}. 
As tube length $l$ increases, $R$ increases.  However, $R$ is not significantly  changed.    
%

%
%





Given the electric charge distribution as shown 
in figure 3,  
we calculated the electric field distribution for multi-rings. 
The  42 rings are used.  
In this case, when $l=4.5$ nm, $R$ is found to be 24 nm. 
We show the electric field line in figure 4.  
In the case of multi-rings, the  curvature of the electric field line  is smooth as  compared with the two-ring case.  
This is because the  complex electric field opposite to the sheath electric field is caused by each ring charge.  

\section{Discussion}
The area $S_0$ through the electric field lines at a sheath edge 
 is also dependent on the charge induced by the external field.  
We set the area to  $S_0(Q, l) $, 
where $Q$ expresses the electric charge induced on the tube.  
According to  Gauss's law, we assume $Q=\alpha E$,
where $\alpha$ denotes the proportional constant related to the polarization. 
%
%
%
%
%
%
%
 %
 From the results of the H$\ddot{\mbox{u}}$ckel-Poisson method,  
 we found that the optimal electric field  
 depends on the tube length as shown in table \ref{table:frontier}.  
 These results will be discussed elsewhere. 
 In this report, to simplify the interpretation, 
 we assume $E^*l=\mit\Delta\phi=$constant, where $E^*$ denotes the optimal electric field 
 to  selectively grow the armchair type. 
 %
 %
We approximated the relation between  the multi-ring charge and the two ring charge as follows. 
\begin{eqnarray}
\label{eq:charge}
Q =Q_s=\alpha E^*=c \frac{\alpha}{l},    
\end{eqnarray}
 where $c$ denotes the constant, $Q_s$ denotes the induced total charge
 summed in the axial direction, 
i.e.  $\int \rho dz$.  
 In the two-ring calculation, $Q$ is regarded as the charge $m e$ in the previous section.  
 
 Under the condition of equation (\ref{eq:charge}), we changed the $\pi$ electron number 
 $m$ corresponding to the induced charge $Q$ and 
 examined the dependence of the interval $R$ on the length $l$. 
First, we set $m=2.3$ and  $l=4.5$ nm. 
Changing $m$ and $l$ under the condition $ml=$constant, 
we obtain the dependence of $R$ on $l$, as shown in table \ref{table:lR}. 
The interval $R$ is not strongly dependent on $l$ and it  weakly decreases.  
Using the conserving law $E_0 S_0=E^*S_1$, where $S_0=\pi R^2(l)$, 
we obtain $E_0(l)=c S_1/(l S_0(l))$, as shown in figure 5.
From the result of figure 5, as the nanotube grows, fortunately, it is found that we should decrease $E_0$  in order to maintain the optimal field at the tip.   
By considering  that $R$ is not strongly dependent on $l$, 
 $E_0$ is approximated by an inverse proportion to the length $l$.

\section{Summary}
The structure of the electric field line out of a nanotube was clarified. 
We found that the potential difference $\mit\Delta\phi$ that gives the optimal electric field to 
grow the armchair-type nanotube is constant for the  tube length. 
The minimum interval $R$ to apply the optimal electric field at the tip was estimated.   
During the initial stage of the growth of the nanotube,  a strong electric field is necessary. 
As  a nanotube grows, the optimal electric field decreases. 
By controlling the sheath electric field  versus time, 
it is possible to  continue to apply the optimal electric field on the nanotube. 


 






\ack
This work was supported by a Grant-in-Aid for The 21st COE Program (Physics of
Self-Organization Systems) at Waseda University from  MEXT.


\newpage
\begin{table}
\caption{Parameters. }
\label{table:1}
\begin{indented}
\item[]\begin{tabular}{@{}ccc} 
			   \br
               physical quantities   & symbols  &    values    \\
\mr
sheath length                          & $L$          & 1 $\mu$m \\
electric field at the sheath edge & $E_0$    & 0.01 V/nm                \\
electric field at the tip of a nanotube & $E_1$    & 1 V/nm($l=$4.5 nm)\\
tube length & $l$                                &   4-500 nm   \\
tube radius & $a$                               &  0.35-1 nm    \\
\br
\end{tabular}
\end{indented}
\end{table}

\Table{The interval $R$ versus the tube length $l$ at the fixed charge,  $m=2.3$. }
\br
length $l$ (nm)&$R$ (nm)&\\
\mr
4.5&19.7&\\
9.0&25&\\
13.5&28&\\
45&37&\\
90&39&\\
450&51&\\
\br
\label{table:R}
\endTable

\begin{table}
\caption{The range of the electric  field 
in which the frontier electron density 
of the armchair type  is greater than that of the zigzag type. 
By using the H$\ddot{\mbox{u}}$ckel-Poisson method, the frontier electron densities at both ends 
of several types of tubes are calculated.   
The optimal value of the  electric field is also shown   
at which  the ratio of the frontier electron density of  
the armchair type to that of  the zigzag type is maximized.}
\label{table:frontier}
\begin{indented}
\item[]\begin{tabular}{@{}ccc} 
			   \br
               tube length $l$ (nm) & electric field range (V/nm) &  optimal electric field $E^*$ (V/nm)     \\
\mr
      4.5 &0.8-1.5& 0.95\\
	        9.0 &0.45-0.7& 0.5\\
			      13.5 &0.4-0.6& 0.45\\
\br
\end{tabular}
\end{indented}
\end{table} 
 
 \begin{table}
\caption{The interval $R$ versus the tube length $l$ under the optimal condition
of equation  (\ref{eq:charge}).  We also changed the $\pi$ electron number $m$. 
Although the interval $R$ decreases, $R$ is not significantly dependent on $l$.  }
\label{table:lR}
\begin{indented}
\item[]\begin{tabular}{@{}cc} 
			   \br
               $l$ (nm)   & R (nm)    \\
\mr
			   4.5 & 19.7\\
			   9.0 & 19.0\\
			   13.5& 18.5\\
			   18.0& 17.0\\
			   22.5& 16.0\\
			   45.0& 13.0\\
\br
\end{tabular}
\end{indented}
\end{table}

\newpage
\begin{figure}
\begin{center}
\epsfxsize=7.5cm
\epsfbox{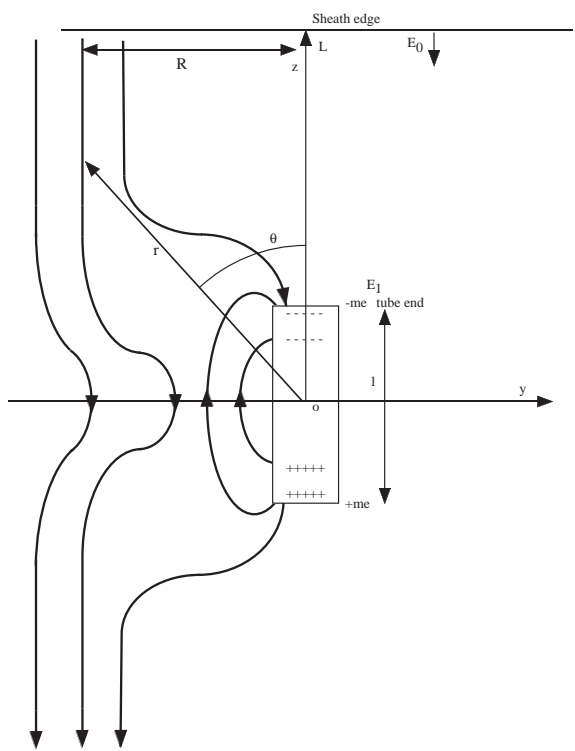}
\label{fig:eline1}
\caption{Outline of the electric field lines. 
The induced charge in the nanotube causes the reversed electric field. 
 The zero field point exists on the $y$ axis when the tube length is short and the external field is weak. }
\end{center}
\end{figure}
\begin{figure}
\begin{center}
\epsfxsize=7.5cm
\epsfbox{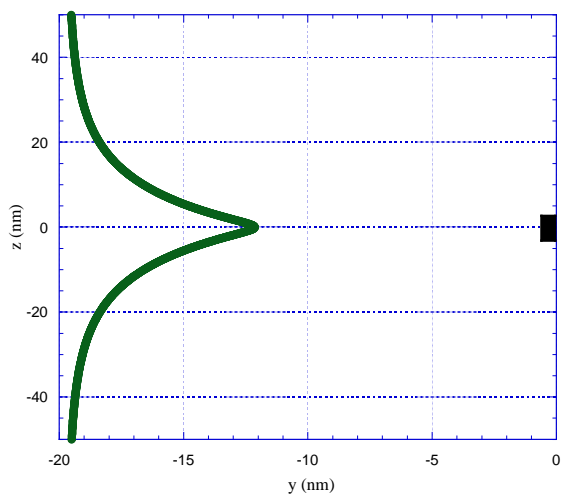}
\caption{The electric field line for two rings. The rectangle
on the right-hand side indicates the  position of the nanotube.}
\end{center}
\label{fig:yztwo}
\end{figure}
\begin{figure}
\begin{center}
\epsfxsize=7.5cm
\epsfbox{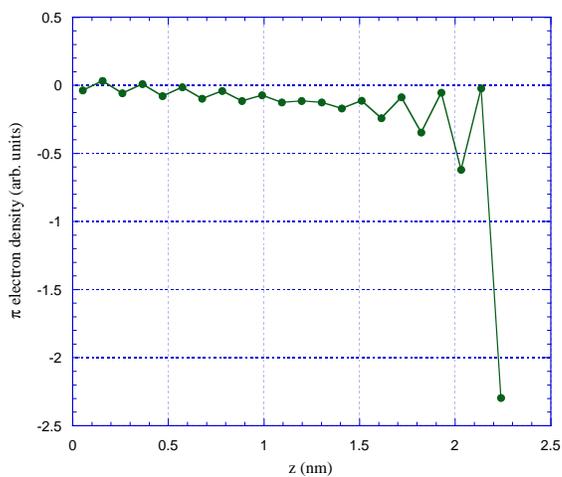}
\caption{A typical $\pi$ electron distribution in the axial position obtained  
using the H$\ddot{\mbox{u}}$ckel-Poisson method. }
\end{center}
\label{fig:pielectron}
\end{figure}
\begin{figure}
\begin{center}
\epsfxsize=7.5cm
\epsfbox{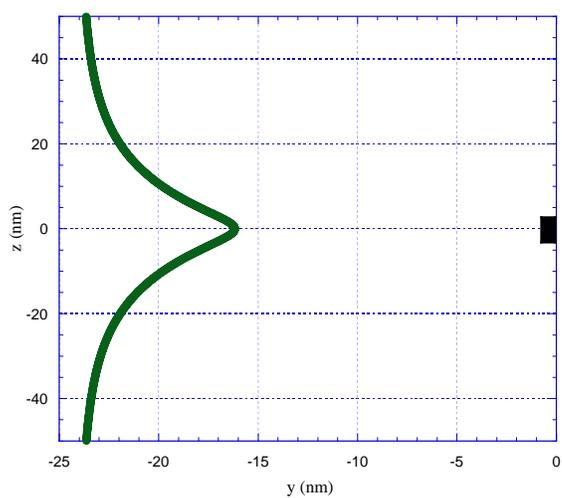}
\caption{The electric field line for multi-rings. In correspondence with the  $\pi$ electron distribution of figure 3,  42 rings are used.  The rectangle
on the right-hand side indicates the  position of the nanotube.}
\end{center}
\label{fig:yzmulti}
\end{figure}
\begin{figure}
\begin{center}
\epsfxsize=7.5cm
\epsfbox{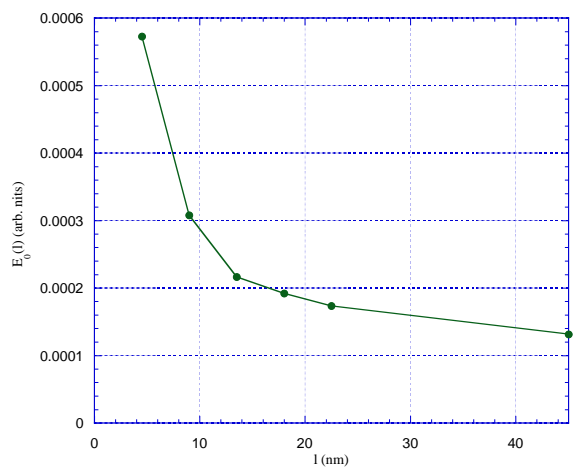}
\caption{The sheath electric field $E_0$ versus the tube length $l$. 
As the  nanotube grows,  
the optimal electric field to grow the armchair type selectively decreases.  
Thus, the sheath field $E_0$ must be decreased with the growth of the tube. }
\end{center}
\label{fig:lR}
\end{figure}

\end{document}